\documentclass[useAMS,usenatbib,usegraphicx]{mn2e}

\usepackage{aas_macros}
\usepackage{amssymb}
\title[A library of high resolution Kurucz spectra]
  {A library of high resolution Kurucz spectra in the wavelength
   range 3000 -- 10000 {\AA}}

\author[T. Murphy and A. Meiksin]
  {T.~Murphy\thanks{Present address: Australia Telescope National Facility, CSIRO, P.O. Box 76, Epping, NSW 1710, Australia}\thanks{Email: Tara.Murphy@csiro.au}, A.~Meiksin \\
  Institute for Astronomy, University of Edinburgh, Royal
  Observatory, Blackford Hill, Edinburgh, EH9 3HJ}
\date{Released 2002 Xxxxx XX}

\pagerange{\pageref{firstpage}--\pageref{lastpage}} \pubyear{2003}

\def\LaTeX{L\kern-.36em\raise.3ex\hbox{a}\kern-.15em
    T\kern-.1667em\lower.7ex\hbox{E}\kern-.125emX}

\def\teff{T_{\rm eff}}

\newcommand{\pegase}{\textsc{pegase}}
\newcommand{\elodie}{\textsc{elodie}}
\newcommand{\synthe}{\textsc{synthe}}
\newcommand{\atlas}{\textsc{atlas9}}

\begin{document}

\label{firstpage}

\maketitle

\begin{abstract}
We present a library of 6410 synthetic spectra with resolution
 $\lambda/\Delta\lambda = 250\,000$ based on the revised Kurucz 1993 model atmospheres.
The library covers the wavelength range $3000 - 10\,000$ {\AA} with 
54 values of effective temperature in the range $5250 - 50\,000$ K,  
11 values of log surface gravity between 0.0 and 5.0 and 
19 metallicities in the range $-5.0$ to 1.0.
We find that, with a few caveats, the library compares well with both the original 20 {\AA} Kurucz spectra and also with observed spectra.
The library is intended for use in population synthesis and physical parameterisation of stellar spectra. 
We assess the suitability of the library for these tasks.
\end{abstract}


\section{Introduction}
Progress in population synthesis and automatic classification of stellar spectra has 
been limited by the spectral resolution of the available synthetic stellar spectra.
The existing synthetic libraries are not at high enough resolution to be useful for 
classifying stars from recent surveys such as the SDSS \citep{stoughton02} ($\lambda/\Delta\lambda\sim1800$), or future surveys 
such as RAVE \citep{steinmetz02} and GAIA \citep{lindegren96}. 
Most classification techniques smooth the observed spectra to the resolution of the synthetic spectra.
This means much of the detailed information in the observed spectra is lost, which 
may reduce the quality of the classifications.
Population synthesis packages such as \pegase\ \citep{fioc97} or
\textsc{gissel} \citep{bruzual93} use a grid of stellar spectra to
generate galaxy spectra.
The resulting galaxy spectra are limited by the resolution of the input stellar spectra.
To study the high resolution features, it is necessary to have a grid of observed or synthetic stellar spectra which match the resolution of the observed galaxy spectra.
For example, galaxy spectra synthesised from 20 {\AA} spectra cannot
be used to measure standard line indices like the Lick indices.

Perhaps the most widely used library of synthetic spectra are the flux
distributions from the Kurucz \atlas\ model atmospheres \citep{kurucz93CD13}. 
It is important to note that while these are usually referred to as
spectra, they are flux distributions predicted directly from the model
atmospheres, rather than spectra generated by a spectral synthesis program. 
The Kurucz atmospheres have several disadvantages which are discussed
in various sources such as \cite{kurucz92}.
However, one of their advantages is the wide range of parameters they
cover, which is important for generating a grid of stellar spectra for 
population synthesis.
The need for higher resolution spectra has been recognised for some time
but, because of the immense computational expense involved, the synthesis of the spectra has been limited to partial wavelength ranges and specific regions of parameter space.
Several groups have generated libraries of spectra from the Kurucz model atmospheres.
For example, \citet*{chavez97} provide a set of 711 Kurucz spectra at $\lambda/\Delta\lambda=250\,000$ in the 
wavelength region $4850 - 5400$ {\AA} and \citet{castelli01} have generated a set of 698 Kurucz 
spectra at $\lambda/\Delta\lambda=20\,000$, in the wavelength region $7650 - 8750$ {\AA} for use with GAIA spectra. 
\citet{gonzalez99} have generated synthetic spectra in very small
spectra regions necessary for a particular application.
They created a grid of synthetic profiles of stellar H Balmer and HeI lines at $\Delta\lambda = 0.3$ {\AA} for the purposes of evolutionary synthesis.

There are also several libraries of observed spectra now available at much higher resolution. For example, the \elodie\ database \citep{prugniel01}, consists of 709 stars observed in the wavelength range $4100-6800$ {\AA} with a resolution of $\lambda/\Delta\lambda \sim 42\,000$.
STELIB \citep{leborgne03} provides spectra for 249 stars observed in the wavelength range $3200-9500$ {\AA} with a resolution of $\lambda/\Delta\lambda \sim2000$.
The observed libraries are crucial for evaluating the accuracy of synthetic spectra and can also be used directly for population synthesis and classification. 
STELIB has been used by \citet{kauffmann03} with a new version of \textsc{gissel}, and \elodie\ has been used to assign physical parameters to stars observed by the SDSS.
However, a limitation of the observed spectral libraries is that they
do not cover the full range in parameter space needed for galaxy population synthesis. 
Complete coverage of the parameter space is even more important for stellar spectral classification.
The most successful approaches to classification have used methods from machine learning 
\citep{bailerjones01a}. 
In these methods, the distribution of spectra in the training set has
a direct impact on the accuracy 
of the classification assigned to new spectra.

There are several efforts to generate higher resolution Kurucz spectra
that are currently in progress.
\citet{bertone02} have generated a grid of 832 spectra at 
$\lambda/\Delta\lambda=500\,000$ over the wavelength range $3500-7000$ {\AA}. 
They intend to extend the wavelength range down to 850 {\AA} at a
resolution of $\lambda/\Delta\lambda=50\,000$. 
\citet*{zwitter03} are in the process of generating a grid of Kurucz
spectra at $\lambda/\Delta\lambda=20\,000$ over the 
wavelength range $2500-10\,500$ {\AA} for use in radial velocity correction work.

We have generated a larger library of 6410 spectra from the Kurucz
model 
atmospheres. 
Previously these spectra were only available either at much lower
resolution (20 {\AA}) or over small wavelength ranges.
Our spectra were generated from \atlas\ model atmospheres, using John
Lester's Unix version of the \synthe\ spectral synthesis package
(Lester 2002, private communication). 
We have modified this package to improve the efficiency of the code, 
making it possible to generate the complete range of Kurucz spectra in
a reasonable time. 

This paper compares our higher resolution spectra with the original 20
{\AA} Kurucz spectra and the STELIB library of observed spectra.
In Section \ref{generate} we describe the main characteristics of the new 
library of spectra.
In Section \ref{comp20} we compare the spectra with the 20 {\AA} Kurucz 
spectra from \citet{kurucz93CD13}.
Finally, in Section \ref{compstellib} we compare the spectra with observed
spectra from the STELIB library.
We will make this library of spectra available for general use on request. 

\section{Generating the Kurucz spectra}\label{generate}
The library presented here has been created using the updated versions of 
the \atlas\ model atmospheres from \cite{kurucz93CD13}.
These are available from Kurucz's website (labelled {\texttt{.dat}). 
Kurucz advises that the \cite{kurucz93CD13} models (labelled \texttt{.datCD}) 
should not be used as they have a discontinuity in the fluxes and colours as a 
function of $\teff$ and $\log(g)$ that was corrected for the revised version.

The models assume plane parallel homogeneous layers in steady-state, 
local thermal equilibrium. 
A microturbulance velocity of 2 ${\rm km s}^{-1}$ and a mixing-length value of 
$\ell/H_p = 1.25$ are used. 
\cite*{castelli97} give a detailed discussion of whether the mixing length theory 
for convection is dealt with adequately in the standard Kurucz atmospheres. 
They have calculated an alternative set of atmospheres in which the Kurucz 
``overshooting'' approximation is not used (NOVER models). 
These models have been shown to predict more accurate observable properties 
(eg. colours) for some atmospheres (\citep{heiter02,smalley97,smalley02}).
However, since they are currently available only for some metallicity values 
(-2.5, -2.0, -1.5, -1.0, -0.5, 0.0, +0.5 dex) we decided not to use these 
atmospheres for population synthesis. 
We have generated a small subset of the NOVER models for comparison with 
the STELIB spectra. 

The adopted atomic line lists for all our spectra are LOWLINES and NLTELINES 
from \cite{kurucz94CD01} and the adopted molecular line list is DIATOMIC 
from \cite{kurucz93CD15}. 
We have included both predicted and measured lines.
The inclusion of the predicted lines is necessary to reproduce accurate flux 
distributions. 
However, it does mean that care should be taken when using the high
resolution spectra, as the properties of individual lines may not be
as accurate as if Kurucz's more up to date linelists (which do not
include predicted lines) were used.  
The molecular lines have been included for spectra with $\teff \le 7000$ K.

The original programs making up the Kurucz package were written in
{\tt FORTRAN77} for the {\tt VAX}. 
John Lester has written a {\tt Unix} version based on the original Kurucz code,
in {\tt FORTRAN77} and a more recent {\tt Unix} version in {\tt
Fortran90}  (Lester 2002, private communication).
However, this code could only be used to generate small sections of a
spectrum at a time because it required massive quantities of disk space.
The complexities in generating spectra with \synthe\ has restricted researchers 
to creating either a small number of spectra, or a large number over a very 
small wavelength range.
We have modified the Lester code to dramatically reduce the disk usage of 
\synthe\ which makes it feasible to generate large spectral ranges on standard 
hardware in a relatively short period of time. 
We have not yet generated spectra for values of $T_{eff} \le 5000$ K. 
These spectra require TiO lines to be included which makes the program 
significantly slower, and requires more disk space and we are investigating 
whether further optimisations may be made.

One thing to note is the subtle difference in terminology between `spectra' and 
`flux distributions' which are sometimes used interchangeably. 
The Kurucz 20 {\AA} flux distributions were predicted directly from the Kurucz
model atmospheres. 
The \atlas\ model atmosphere program sums up the opacity in broad 10 or 20 
{\AA} bins creating a low resolution spectrum. 
The Kurucz spectra that we have generated have been calculated from the model 
atmospheres, using the \synthe\ software. 
These spectra are generated at extremely high resolution (ideally as high as 
the computational resources allow) and then rebinned to the required resolution.
The newest versions of \atlas\ will combine both the model atmosphere 
generation and the spectral synthesis into one process. 

Of the 7216 model atmospheres available in Kurucz's standard distribution, 
we have generated a grid of 6410 spectra, which excludes the lowest temperature 
spectra.
The spectra cover the wavelength range $3000-10\,000$ {\AA} which was chosen 
to be useful for comparisons with SDSS spectra.
The spectra were generated at a resolution of $\lambda/\Delta\lambda = 250\,000$.
\begin{figure}
\includegraphics[height=84mm, angle=90]{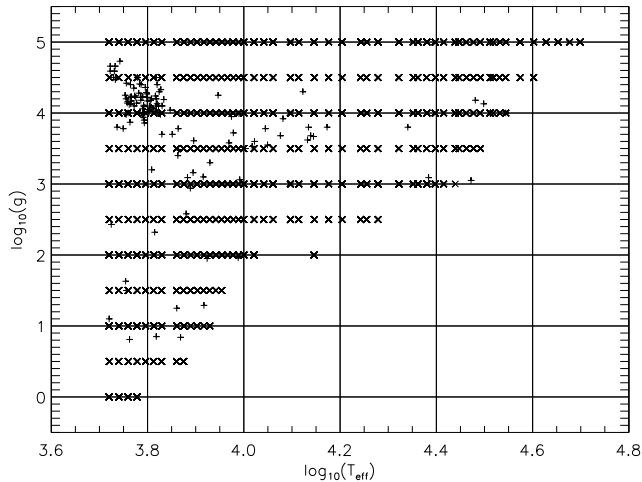}
\caption{The distribution of our Kurucz spectra in $\log(\teff)$ -
$\log(g)$ space ($\times$). Also plotted is the distribution of the
STELIB spectra with $\teff > 5000$ K ($+$).}
\end{figure}

\section{Comparisons with the 20{\AA} Kurucz spectra} \label{comp20}
In this section we compare our library of spectra with the revised version 
of the 20 {\AA} Kurucz flux distributions.
There have been several changes to the software and data between the
original release of the 20 {\AA} Kurucz flux distributions, and the present. 
Also, the flux distributions are predicted directly from the model atmosphere 
code, so it is not possible to compare them directly with the spectra generated by
spectral synthesis.
Because of this, the spectra we have generated are not expected to match the flux 
distributions exactly, but it is an important check to make sure there was 
broad agreement between the spectra. 

\subsection{Direct comparisons} \label{dc1}
We compare the higher resolution Kurucz spectra with
the 20 {\AA} flux distributions. 
Figure \ref{20aspec} shows several examples of these comparisons.
For the purposes of this comparison, we re-binned the new spectra
to 20 {\AA} using a simple top-hat function. 
\begin{figure*}
\includegraphics[width=168mm]{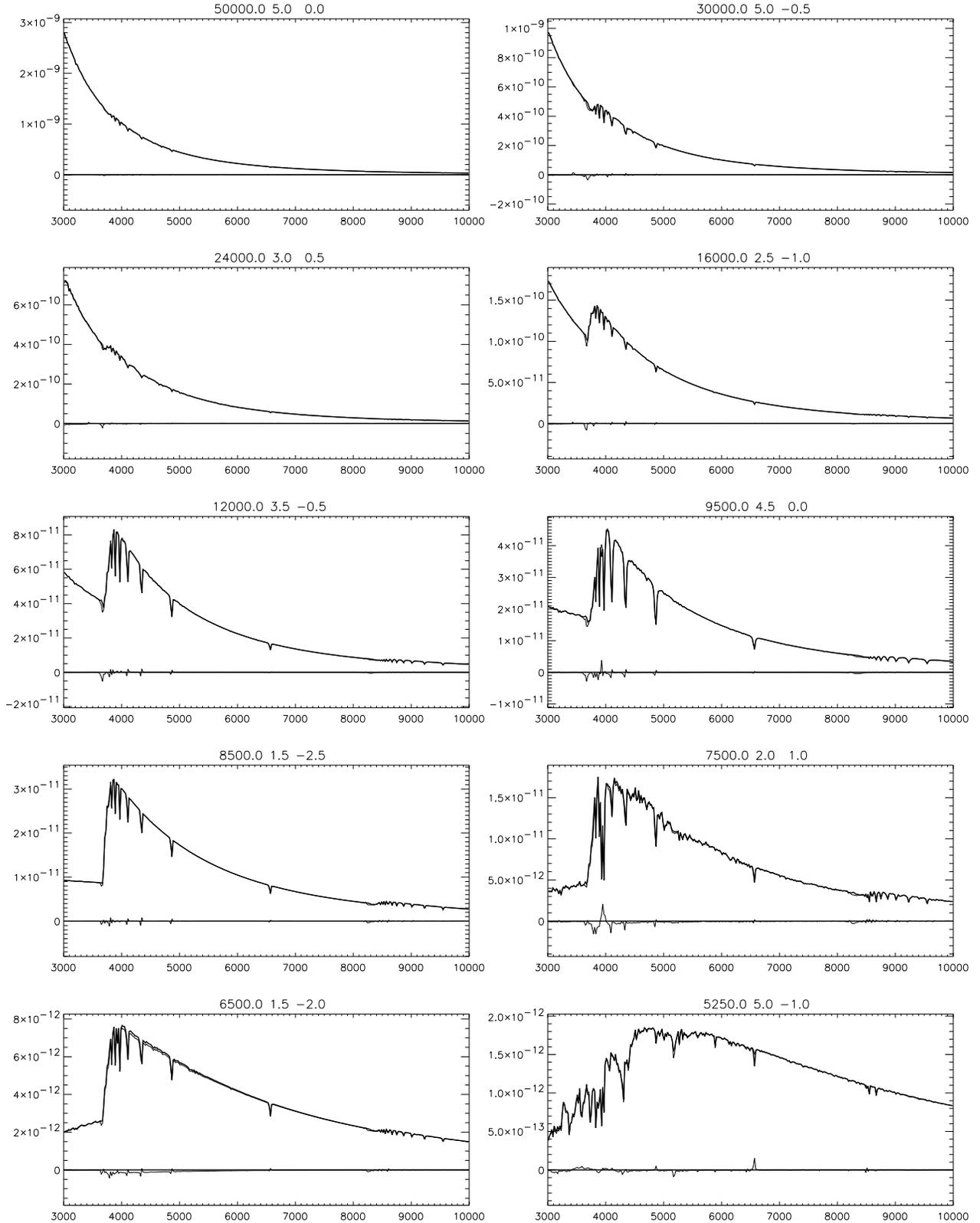}
\caption{Comparison between a sample of the 20 {\AA} Kurucz flux distributions 
(bold line) and the newly generated Kurucz spectra (thin line) (box-car smoothed 
to $\Delta\lambda = 20${\AA}). These spectra cover a range of physical parameters: 
the title of each plot gives the values for $\teff$, $\log(g)$ and $[Fe/H]$. 
The $x$-axis is the wavelength in {\AA} and the $y$-axis is flux, $F(\lambda)$, 
in arbitrary units. 
Underneath each pair of spectra is 
$\Delta F = F(\lambda)_{\rm new} - F(\lambda)_{\rm old}$. \label{20aspec}}
\end{figure*}
There are several differences between the two sets of spectra:
\begin{enumerate}
\item In all the spectra there is an extra dip in the new spectra at around 
3700 {\AA}. 
This is due to inadequate treatment of the Balmer jump. 
In the model atmosphere code, the way the Balmer jump is dealt with
has been modified in order to remove the dip (Kurucz, private communication).
\item In some of the lower temperature spectra, the shape of the Ca H/K 
doublet is not identical. 
This is largely due to the fact that for the model atmospheres, the
Stark broadening was artificially increased to make up for missing
lines in computing the distribution function (Kurucz, private communication). 
This was changed back to generate the synthetic spectra, and also for more 
recent flux distributions. 
\item In lower temperature spectra ($\teff \le 7000$ K) there is a systematic 
difference in flux in the G-band. This is likely to be due to changes in the 
molecular line lists since the flux distributions were produced. 
\item Most of the strong lines have slightly different depths. These small peaks 
are likely to be unresolved line cores, caused by the fact that the 20 {\AA} flux 
distributions are undersampled, whereas our new synthetic spectra are generated at 
high resolution. 
\end{enumerate}
Probably the most significant of these points for our applications is the 
difference in the G-band flux.
This affects individual lines, and the $U$ band magnitude of the spectra.
In the following sections we quantify this. 

\subsection{Colours}
Comparing the $UBVRI$ magnitudes is important for checking that the 
broadband properties of the spectra are reproduced.
We have used the Johnson-Cousins bandpasses, as defined by
\cite{bessell90}.
Figure \ref{kkcol} shows the offset in magnitudes calculated for the new
spectra and the 20 {\AA} flux distributions.
The RMS errors for the colours for the two sets of spectra are: 
$\Delta U = 0.05$ mag, $\Delta B = 0.01$ mag, $\Delta V = 0.01$ mag, 
$\Delta R = 0.01$ mag and $\Delta I = 0.01$ mag.
The $B$, $V$, $R$ and $I$ magnitudes are in good agreement.
However, the differences are greater for the $U$ magnitudes at lower
temperatures ($\teff \le 7000$ K). 
For higher temperature spectra the $U$ magnitudes are also in good agreement.
The $U$ band discrepancy is probably a consequence of the differences in 
the spectra noted in Section \ref{dc1}.
\begin{figure*}
\includegraphics[height=168mm, angle=90]{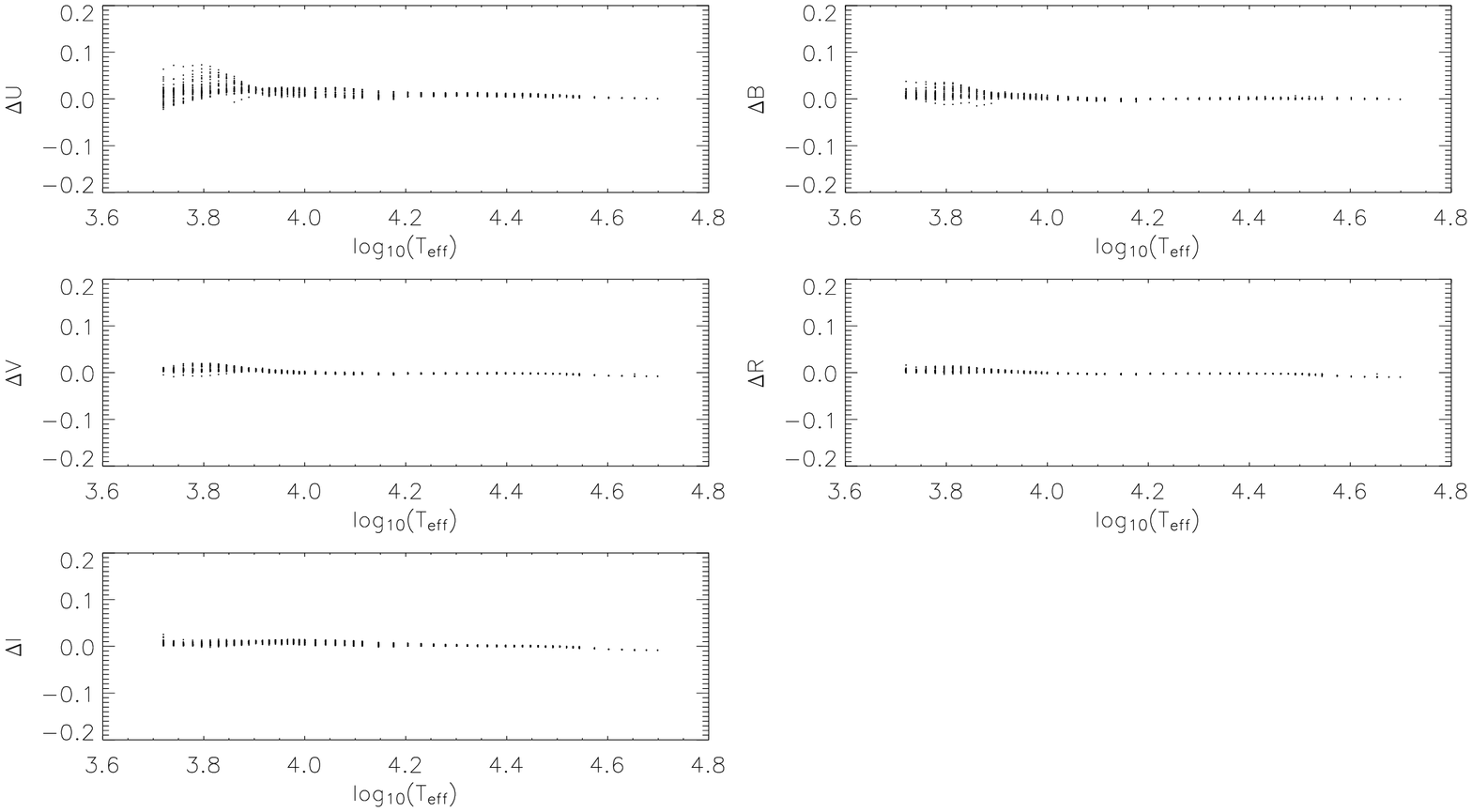}
\caption{Magnitude offsets between our Kurucz spectra and the 20 {\AA}
flux distributions. The x-axis is $log_{10}(T_{eff})$ and the y-axis
shows the difference (in magnitudes) between the Kurucz spectra we
have generated and the 20 {\AA} flux distributions. \label{kkcol}}
\end{figure*}

The question is how important these magnitude offsets are for the
applications we are interested in.
A difference in measured colour is largely degenerate with a difference in
$\teff$. 
For example, \cite*{lejeune97} find that the colour vs. temperature
relations for Kurucz spectra do not match those derived empirically.
To fix the problem, they developed a method for adjusting the shape of
the continuum (in other words change the magnitudes) for spectra of a
certain $\teff$.
However, an alternative approach is to leave the shape of the spectrum
unchanged and reassign a $\teff$ to the spectrum.
For example, a solar metallicity spectrum with $\teff =
7500$ and $\log(g) = 1.0$ has a $U$ magnitude offset of 0.05.
This difference in colour between our new 
spectrum and the Kurucz flux distribution corresponds to a change of
$\teff$ by less than 30 K. 
Since quoted errors for temperatures on observed spectra are typically
around $40 - 100$ K (see for example \cite{katz98} or
\cite*{alonso99}), the spectra agree to within usual measurement errors.

\section{Comparisons with observed spectra}\label{compstellib}
Having shown that the new spectra agree reasonably well with the 20
{\AA} flux distributions, the next step is to evaluate them against observed spectra.
This serves two purposes. 
Firstly, there are some differences between the two sets of Kurucz
spectra, especially in the blue. However, if these differences are
smaller than the differences between one of the Kurucz spectra and an
observed spectrum with the same physical parameters, then they may be neglected.
Secondly, we want to evaluate what kind of applications the synthetic
spectra can be used for.
Our aim is to demonstrate that our new Kurucz spectra are as suitable
for use in population synthesis as the commonly used 20 {\AA} Kurucz
flux distributions, but with the advantage of higher resolution.
There are already various comparisons of Kurucz spectra with observed
spectra in the literature.
For example, the colour$-\teff$ correlation for the spectra has been
analysed by \citet{lejeune97} and the 20 {\AA} spectra have been
compared with observed spectra by \citet*{straizys97} and \citet*{straizys02}.
We here make our own comparisons to focus on properties such as the
Lick indices which are relevant to the specific applications we are interested in.

We have compared the spectra to those in the STELIB library \citep{leborgne03},
which consists of 250 spectra in the visible range ($3200 - 9500$
{\AA}) with a spectral resolution of $\Delta\lambda \lesssim 3$ {\AA}
and covering a wide range in parameter space.
As well as a direct comparison of the SEDs, we compare the measured
colours and Lick indices to confirm that the spectra meet out requirements.

Many of the STELIB stars do not have values for all three physical
parameters ($\teff$, $\log(g)$ and $[Fe/H]$), have large section of
the spectrum missing, or do not lie close to a point on our grid of spectra.
Hence we have selected a subset of 125 of the spectra that have
tabulated values for $\teff$, $\log(g)$ and $[Fe/H]$ in
\citet{leborgne03}, $\teff > 5000K$, and which have a good $\chi^2$
match to the corresponding Kurucz spectrum.
The values for most of the physical parameters were obtained from the 
\citet{destrobel97} and \citet*{destrobel01} catalogues. 
However, some were obtained from the \elodie\ database
\citep{prugniel01} and several were calculated using the \textsc{tgmet}
method \citep{katz98}. 
We corrected the spectra for interstellar reddening using the values of
$A_v$ from \citet{leborgne03} and the \textsc{idl astrolib} routine 
\textsc{ccm\_dered.pro}. 
The STELIB spectra have not been corrected for atmospheric absorption,
so they have strong absorption features (most noticeably in the Oxygen
A-band around 7600{\AA}) which are not present in the synthetic spectra. 

\subsection{Colours}
We have calculated colours for all of the STELIB spectra
and their closest Kurucz matches.
Since the wavelength range of the STELIB spectra is $3200 - 9500$
{\AA}, we 
have calculated a modified $U$ band magnitude (denoted $U^*$), which
is truncated at 
$3200$ {\AA} for both the STELIB and Kurucz spectra. 
Figure \ref{cc2} shows the magnitude offsets between the STELIB
spectra and their Kurucz matches. 
The RMS error between the two sets of spectra are: $\Delta U^*=0.14$
mag, $\Delta B=0.05$ mag, $\Delta V=0.02$ mag, $\Delta R=0.03$ mag and
$\Delta I=0.04$ mag.
For comparison, we also calculated the magnitude offsets for the
STELIB spectra and their closest Kurucz matches, using the 20 {\AA}
flux distributions.
The RMS error between these two sets of spectra are: $\Delta U^*=0.11$
mag, $\Delta B=0.05$ mag, $\Delta V=0.02$ mag, $\Delta R=0.03$ mag and
$\Delta I=0.04$ mag.
This demonstrates that, with the exception of $U^*$, the scatter in the
magnitudes is no worse for our new spectra than for the 20 {\AA}
spectra that are in standard use. 

Some differences are expected due to the mismatch between the STELIB
temperatures and those of the closest Kurucz match -- many of the
spectra have a difference in temperature of about 100 K, corresponding
to a difference of a few hundredths in magnitude. 
However, the important point for our work is that our new spectra are
as well matched to the observed spectra as the 20 {\AA} spectra that
are often used for population synthesis.
\begin{figure*}
\includegraphics[height=168mm, angle=90]{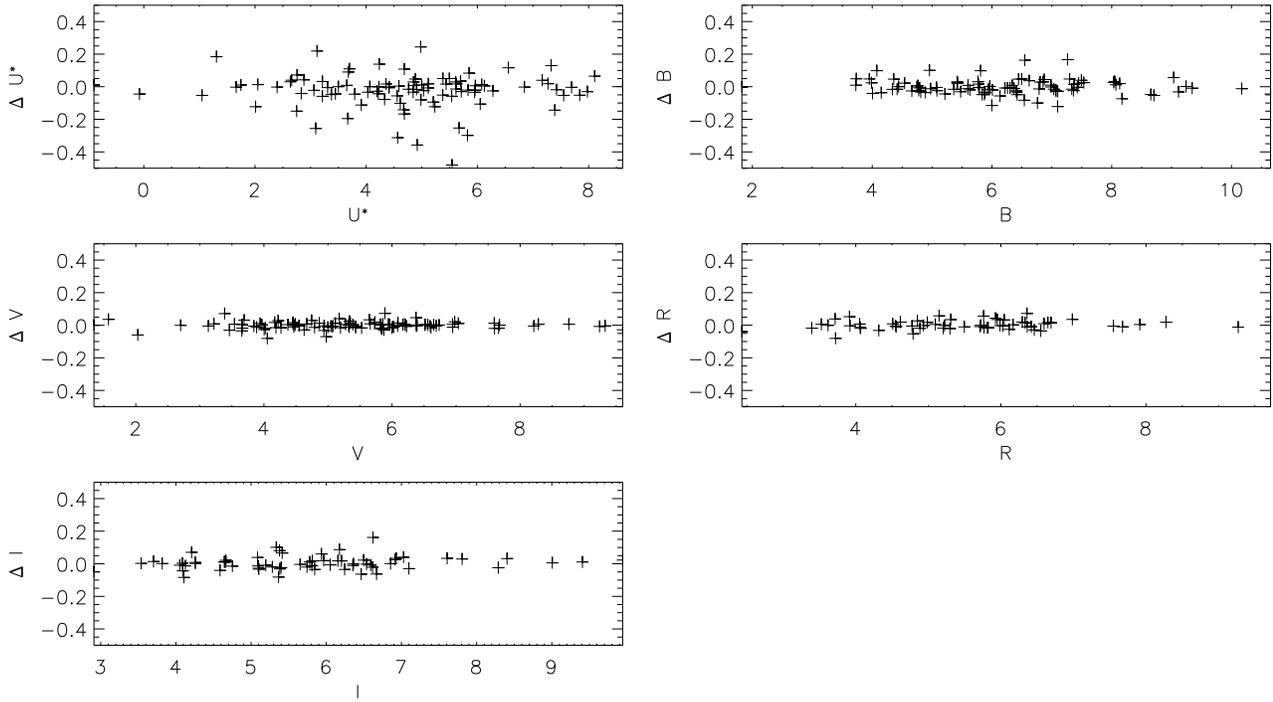}
\caption{Magnitude offsets between the STELIB spectra and the closest
Kurucz spectrum match from the grid.\label{cc2}}
\end{figure*}
Figure \ref{cc} shows various colour comparisons.  
There is some scatter in each of the plots and a slight systematic 
offset in $R-I$.
Again however, this is no worse than when comparing the 20 {\AA} 
spectra with the STELIB spectra.
\begin{figure*}
\includegraphics[height=168mm, angle=90]{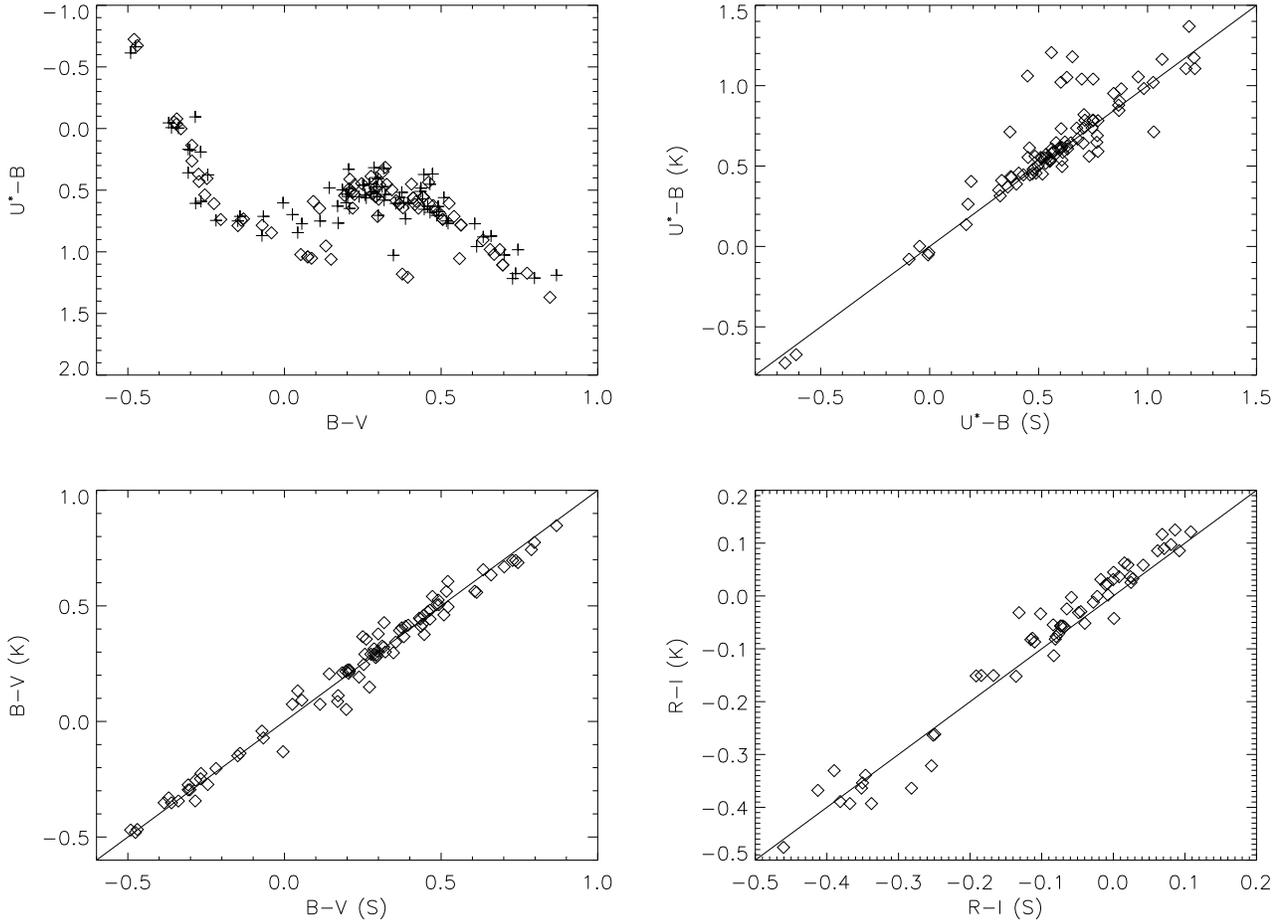}
\caption{Top Left: $U^*-B$ vs. $B-V$ for STELIB spectra
($+$) and their Kurucz matches ($\diamond$). Top Right: $U^*-B$ colours for STELIB
spectra ($x$-axis) and their Kurucz matches ($y$-axis). The
solid line is the line ``$x=y$'' for comparison. Bottom Left: Comparison of 
$B-V$ colours. Bottom Right: Comparison of $R-I$ colours.  \label{cc}}
\end{figure*}

\subsection{Line indices}
We have calculated Lick indices for each of the 125 STELIB spectra in
our sample, and for the matching Kurucz spectra.
There are two types of index, atomic ($I_a$) and molecular ($I_m$)
which we have calculated from the standard formulae:
\begin{equation}
I_a \equiv \int_{\lambda_{c1}}^{\lambda_{c2}} \left[1 - \frac{S(\lambda)}{C(\lambda)}\right] d\lambda
\end{equation}
\begin{equation}
I_m \equiv -2.5\log_{10} \frac{\int_{\lambda_{c1}}^{\lambda_{c2}}\frac{S(\lambda)}{C(\lambda)} d\lambda}{\lambda_{c2} - \lambda_{c1}}
\end{equation}
where $\lambda_{c1}$ and $\lambda_{c2}$ are the limits of the central 
bandpass defining the index (in {\AA}), $S(\lambda)$ is the object
spectrum and $C(\lambda)$ is the linearly interpolated pseudo-continuum, defined by
\begin{equation}
C(\lambda) \equiv S_b \frac{\lambda_r - \lambda}{\lambda_r - \lambda_b} + 
                  S_r \frac{\lambda - \lambda_b}{\lambda_r -
                  \lambda_b} .
\end{equation} 
Here,
\begin{equation}
S_b \equiv \frac{\int_{\lambda_{b1}}^{\lambda_{b2}}S(\lambda) d\lambda}
                {\lambda_{b2} - \lambda_{b1}}, \quad
\lambda_b \equiv (\lambda_{b1} + \lambda_{b2})/2
\end{equation}
\begin{equation}
S_r \equiv \frac{\int_{\lambda_{r1}}^{\lambda_{r2}}S(\lambda) d\lambda}
                {\lambda_{r2} - \lambda_{r1}}, \quad
\lambda_r \equiv (\lambda_{r1} + \lambda_{r2})/2
\end{equation}
and $\lambda_{b1}$, $\lambda_{b2}$, $\lambda_{r1}$ and $\lambda_{r2}$
are the limits of the blue and red continuum bands.
The bandpass definitions for the Lick indices and the red and blue 
pseudo-continua are those defined in \citet{worthey94}. 
These are given in Table \ref{bandpass}.

A method for random error estimation in line-strength indices is
outlined in detail by \citet{cardiel98}. 
The resulting equations are based on a full analysis of the error
propagation throughout the calculation process.
The resulting random errors are given by:
\begin{eqnarray}\label{iaerr}
\sigma^2[I_a] = \sum^{N_{\rm pixels}}_{i=1}\left[
	\frac{C^2(\lambda_i)\sigma^2(\lambda_i) + 
	S^2(\lambda_i)\sigma^2_{C(\lambda_i)}}{C^4(\lambda_i)}\right] 
	d\lambda^2_i + \nonumber \\
	\sum^{N_{\rm pixels}}_{i=1}\sum^{N_{\rm pixels}}_{j=1, j\ne i}\left[ 
	\frac{S(\lambda_i)S(\lambda_j)}{C^2(\lambda_i)C^2(\lambda_j)}
	(\Lambda_1\sigma^2_{S_b} + \Lambda_4\sigma^2_{S_r})\right]
	d\lambda_i d\lambda_j
\end{eqnarray}
for the atomic indices and:
\begin{equation}\label{imerr}
\sigma[I_m] = 2.5 \frac{\log(e)}{10^{-0.4I_m}} \frac{1}{\lambda_{c_2} - \lambda_{c_1}} \sigma[I_a]
\end{equation}
for the molecular indices, where
\begin{equation}
\Lambda_1 \equiv \frac{(\lambda_r - \lambda_i)(\lambda_r - \lambda_j)}
	{(\lambda_r - \lambda_b)^2}
\end{equation}
\begin{equation}
\Lambda_4 \equiv \frac{(\lambda_i - \lambda_b)(\lambda_j - \lambda_b)}
	{(\lambda_r - \lambda_b)^2}
\end{equation}
\begin{equation}
\sigma^2_{C(\lambda_i)} = 
	\left(\frac{\lambda_r - \lambda_i}{\lambda_r - \lambda_b}\right)
	\sigma^2_{S_b} + 
	\left(\frac{\lambda_i - \lambda_b}{\lambda_r - \lambda_b}\right)
	\sigma^2_{S_r} 
\end{equation}
\begin{equation}
\sigma^2_{S_b} = \frac{1}{(\lambda_{b_2} - \lambda_{b_1})^2}
	\sum^{N_{\rm pixels(blue)}}_{i=1} \sigma^2(\lambda_i) \; d\lambda_i^2
\end{equation}
\begin{equation}
\sigma^2_{S_r} = \frac{1}{(\lambda_{r_2} - \lambda_{r_1})^2}
	\sum^{N_{\rm pixels(red)}}_{i=1} \sigma^2(\lambda_i) \;
	d\lambda_i^2 \; .
\end{equation}
Note that \citet{cardiel98} assume the size of
each pixel $d\lambda$ ($\Theta$ in their notation) is fixed, and so
take it out of all the summations. 
However, in general (and in our case), $d\lambda$ is not fixed, and so
we keep it inside the summation signs. 
\begin{table*}
\begin{center}
\begin{tabular}{l|cccc}\hline\hline
Name & Central Bandpass & Blue Continuum & Red Continuum & Type \\
\hline\hline
$\mbox{CN}_1$ &4142.125 -- 4177.125 & 4080.125 -- 4117.625 & 4244.125 -- 4284.125 & M \\
$\mbox{CN}_2$ &4142.125 -- 4177.125 & 4083.875 -- 4096.375 & 4244.125 -- 4284.125 & M \\
$\mbox{Ca}4227$ &4222.250 -- 4234.750 & 4211.000 -- 4219.750 & 4241.000 -- 4251.000 & A \\
$\mbox{G}4300$ &4281.375 -- 4316.375 & 4266.375 -- 4282.625 & 4318.875 -- 4335.125 & A \\
$\mbox{Fe}4383$ &4369.125 -- 4420.375 & 4359.125 -- 4370.375 & 4442.875 -- 4455.375 & A  \\
$\mbox{Ca}4455$ &4452.125 -- 4474.625 & 4445.875 -- 4454.625 & 4477.125 -- 4492.125 & A \\
$\mbox{Fe}4531$ &4514.250 -- 4559.250 & 4504.250 -- 4514.250 & 4560.500 -- 4579.250 & A \\
$\mbox{Fe}4668$ &4634.000 -- 4720.250 & 4611.500 -- 4630.250 & 4742.750 -- 4756.500 & A \\
$\mbox{H}_\beta$ &4847.875 -- 4876.625 & 4827.875 -- 4847.875 & 4876.625 -- 4891.625 & A\\
$\mbox{Fe}5015$ &4977.750 -- 5054.000 & 4946.500 -- 4977.750 & 5054.000 -- 5065.250 & A \\
$\mbox{Mg}_1$ &5069.125 -- 5134.125 & 4895.125 -- 4957.625 & 5301.125 -- 5366.125 & M \\
$\mbox{Mg}_2$ &5154.125 -- 5196.625 & 4895.125 -- 4957.625 & 5301.125 -- 5366.125 & M \\
$\mbox{Mg}_b$ &5160.125 -- 5192.625 & 5142.625 -- 5161.375 & 5191.375 -- 5206.375 & A  \\
$\mbox{Fe}5270$ &5245.650 -- 5285.650 & 5233.150 -- 5248.150 & 5285.650 -- 5318.150 & A \\
$\mbox{Fe}5335$ &5312.125 -- 5352.125 & 5304.625 -- 5315.875 & 5353.375 -- 5363.375 & A \\
$\mbox{Fe}5406$ &5387.500 -- 5415.000 & 5376.250 -- 5387.500 & 5415.000 -- 5425.000 & A \\
$\mbox{Fe}5709$ &5696.625 -- 5720.375 & 5672.875 -- 5696.625 & 5722.875 -- 5736.625 & A \\
$\mbox{Fe}5782$ &5776.625 -- 5796.625 & 5765.375 -- 5775.375 & 5797.875 -- 5811.625 & A \\
$\mbox{Na}_D$ &5876.875 -- 5909.375 & 5860.625 -- 5875.625 & 5922.125 -- 5948.125 & A \\
\hline\hline
\end{tabular}
\caption{\label{bandpass}Bandpass definations for the Lick indices
calculated here. Each index has a type A or M signifying atomic and
molecular indices, respectively. Atomic indices are measured in {\AA}
and molecular indices are measured in magnitudes. Note that these are
air wavelengths. }
\end{center}
\end{table*}

Figure \ref{lickcompare} shows the correlation between the Lick
indices calculated from the STELIB spectra and the Kurucz spectra.
We have not calculated the two TiO indices since TiO lines are switched
off in these spectra.
The average absolute offsets (ie: $\sum \frac{|I_k - I_s|}{n}$) for
the correlations are given in Table \ref{lickerr}.
It should be noted that the Kurucz spectrum paired with a given
STELIB spectrum is not necessarily the best possible spectrum that
could be synthesised to match the STELIB spectrum, but simply the 
spectrum from our grid with the closest match in physical parameters. 
This means it does not take into account any particular properties 
of the STELIB spectrum that could in principle be modelled and which may 
affect the line indices.
However, the comparison does give an indication of the level of
mismatch to be expected when comparing observed stellar spectra with a
grid of synthetic spectra like ours.
Such mismatches are an inevitable consequence of any automated
comparisons between a grid of model spectra and the vast number of
spectra being made available from large surveys like the SDSS. 
\begin{figure*}
\includegraphics[width=168mm]{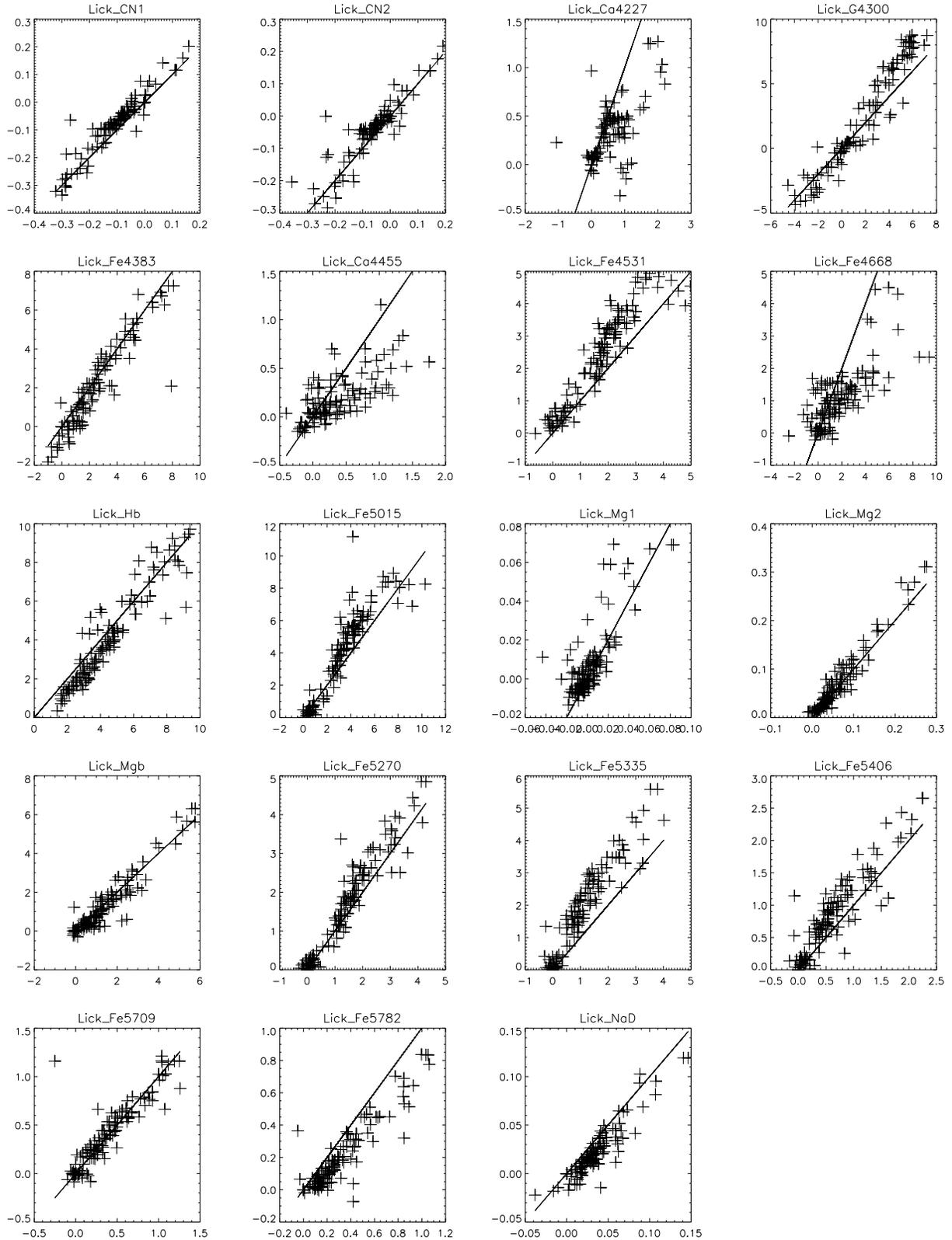}
\caption{Each plot shows the correlation between the Lick index
calculated on the STELIB spectrum ($x$-axis) and the closest matching
Kurucz spectrum from our grid ($y$-axis). The solid line is the line
``$x=y$'' for comparison. We have not included the TiO indices for
reasons discussed in the text. The RMS error for each index is given
in Table \protect\ref{bandpass}. \label{lickcompare}} 
\end{figure*}
\begin{table}
\begin{center}
\begin{tabular}{lcccc}\hline\hline
Name & Mean value & Av. Abs. Err. & Med. Abs. Err. & Type \\
\hline\hline
$\mbox{CN}_1$ & -0.107 & 0.03 & 0.02 & M \\
$\mbox{CN}_2$ & -0.066 & 0.03 & 0.01 & M \\
$\mbox{Ca}4227$ & 0.583 & 0.33 & 0.03 & A \\ 
$\mbox{G}4300$ & 2.028 & 1.28 & 0.42 & A \\ 
$\mbox{Fe}4383$ & 2.593 & 0.63 & 0.95 & A \\ 
$\mbox{Ca}4455$ & 0.366 & 0.27 & 0.27 & A \\ 
$\mbox{Fe}4531$ & 1.698 & 0.72 & 0.02 & A \\ 
$\mbox{Fe}4668$ & 1.732 & 1.12 & 2.17 & A \\ 
$\mbox{H}_\beta$ & 4.430 & 0.76 & 0.66 & A \\ 
$\mbox{Fe}5015$ & 3.342 & 0.96 & 0.08 & A \\ 
$\mbox{Mg}_1$ & 0.003 & 0.01 & 0.00 & M \\
$\mbox{Mg}_2$ & 0.061 & 0.01 & 0.02 & M \\
$\mbox{Mg}_b$ & 1.430 & 0.33 & 0.07 & A \\ 
$\mbox{Fe}5270$ & 1.493 & 0.34 & 0.02 & A \\ 
$\mbox{Fe}5335$ & 1.123 & 0.82 & 0.18 & A \\ 
$\mbox{Fe}5406$ & 0.640 & 0.27 & 0.58 & A \\ 
$\mbox{Fe}5709$ & 0.377 & 0.09 & 0.26 & A \\ 
$\mbox{Fe}5782$ & 0.323 & 0.13 & 0.49 & A \\ 
$\mbox{Na}_D$ & 0.036 & 0.01 & 0.06 & M \\
\hline\hline
\end{tabular}
\caption{\label{lickerr} Statistical comparison between the STELIB
Lick indices and those of their Kurucz matches. Column 2 shows the
mean value of the STELIB index. Column 3 shows the average absolute
offset between the Kurucz and STELIB indices. Column 4 shows the
median absolute offset between the Kurucz and STELIB indices.}
\end{center}
\end{table}

We have calculated Lick indices and their associated errors for a set
of 9473 SDSS galaxies from the Early Data Release \citep{stoughton02}.
This allows us to compare the scatter between the Lick index values
for the STELIB and Kurucz spectra, with the expected accuracy of the
indices measured from observed galaxy spectra.  
We have found that the random error (as calculated by Equations
\ref{iaerr} and \ref{imerr}) associated with each index measured in
the SDSS spectra, is of the same order of magnitude as the average
offset between the Kurucz and STELIB index measurement. 
This suggests that the indices measured from galaxy spectra synthesized from
the Kurucz spectra will have at least comparable accuracy to those measured with
SDSS, and so our grid of stellar spectra is well matched to the
SDSS galaxies.

\section{Discussion}
We have generated a grid of theoretical spectra from the Kurucz model 
atmospheres. 
Since the intended use of these spectra is in population synthesis and
stellar classification, we have made several comparisons to check the
validity of using the spectra for these purposes.
The broadband properties of the spectra compare well with observed spectra, 
as do the line index measurements.
The comparisons do not guarantee the accuracy of the Kurucz spectra as
models of observed stellar spectra.
Rather, they demonstrate that our new, high resolution Kurucz spectra
are as good for use in population synthesis as the commonly used
library of 20 {\AA} spectra.
The advantage of these spectra over the previously available Kurucz
spectra is that they allow the modelling of spectral line features
such as the Lick indices.

When using these spectra at high resolutions, it should be recognised that 
the spectra were generated using line lists that include `predicted' lines. 
This is necessary to reproduce the broadband colours of the spectra accurately. 
However, it does mean that many of the individual lines present at
high resolutions do not have measured properties. 
Also, the line lists that we used (such as LOWLINES) are known to have problems 
with the values for specific lines.
As this mostly involves weak lines, it should not present any
difficulties for population synthesis at the resolution of the SDSS.
More accurate properties for these lines could be obtained using
alternative values, for example from the Vienna Atomic Line Database 
(VALD) \citep{kupka00}. 

Also, the 20 {\AA} Kurucz flux distributions are often used with the
corrections of \cite{lejeune97,lejeune98} applied. 
The corrections are an attempt to
calibrate the spectra by comparing the synthetic model colours
with empirical stellar colours. We investigated applying the same
corrections to our high resolution spectra, but found that artifacts were 
introduced when using this technique directly.
The Lejeune corrections mainly affect the lower 
temperature spectra, which we have not included in this library. 
For most of our spectra, the corrections are negligibly small.

We intend to extend the library to the lower temperature models in which TiO 
lines become important and also generate spectra for the NOVER models as these 
are more accurate. 
We are in the process of building a large library for population synthesis -- 
a higher resolution version of that done by \cite{lejeune97} -- which will also
incorporate the NextGen models \citep*{hauschildt99a}, \citep{hauschildt99b}.

\section*{Acknowledgements}
TM thanks the University of Edinburgh and the University of Sydney for scholarships. 
We would like to thank John Lester for making his code available and
for the extensive help he provided in generating the spectra.
We would also like to thank Ivan Baldry, Friedrich Kupka, Barry
Smalley, Robert Kurucz and Ian Sheret for helpful discussions and
email exchanges about various aspects of the spectra and James Curran
for advice on computational issues.

\bibliographystyle{mn}



\makeatletter
\def\thebiblio#1{%
 \list{}{\usecounter{dummy}%
         \labelwidth\z@
         \leftmargin 1.5em
         \itemsep \z@
         \itemindent-\leftmargin}
 \reset@font\small
 \parindent\z@
 \parskip\z@ plus .1pt\relax
 \def\newblock{\hskip .11em plus .33em minus .07em}
 \sloppy\clubpenalty4000\widowpenalty4000
 \sfcode`\.=1000\relax
}
\let\endthebiblio=\endlist
\makeatother

\label{here}



\label{lastpage}

\end{document}